\documentclass[conference]{IEEEtran}

\pdfoutput=1

\usepackage[colorlinks=true,linkcolor=black,citecolor=black,filecolor=black,urlcolor=black,plainpages=false,pdfpagelabels,breaklinks=true]{hyperref}

\usepackage[utf8]{inputenc}
\usepackage{algorithmic}
\usepackage{amsthm,amsmath}
\usepackage{amsfonts, amssymb, amscd}
\usepackage{bbm}
\usepackage[compress]{cite}
\usepackage[capitalize]{cleveref}
\usepackage{glossaries}
\usepackage{graphicx}
\usepackage{here}
\usepackage{latexsym}
\usepackage[activate={true,nocompatibility},final,tracking=true,kerning=true,spacing=true,factor=1100,stretch=10,shrink=10]{microtype}
\usepackage[binary-units]{siunitx}
\usepackage[disable]{todonotes}

\newcommand{\todoin}[1]{\todo[inline]{#1}}

\newacronym{ppu}{PPU}{plasticity processing unit}
\newacronym{padi}{PADI}{parallel driver interface}

\newacronym{sram}{SRAM}{static random-access memory}
\newacronym{adc}{ADC}{analog-to-digital converter}
\newacronym{cadc}{CADC}{column-parallel analog-to-digital converter}
\newacronym{dac}{DAC}{digital-to-analog converter}
\newacronym{dut}{DUT}{design under testing}
\newacronym{asic}{ASIC}{application-specific integrated circuit}
\newacronym{fpga}{FPGA}{field-programmable gate array}
\newacronym{vlsi}{VLSI}{very-large-scale integration}
\newacronym{simd}{SIMD}{single instruction, multiple data}
\newacronym{ocp}{OCP}{Open Core Protocol}
\newacronym{soc}{SoC}{system on a chip}
\newacronym{sta}{STA}{static timing analysis}
\newacronym{gals}{GALS}{globally asynchronous locally synchronous}
\newacronym{dpi}{DPI}{SystemVerilog direct programming interface}
\newacronym{pll}{PLL}{phase-locked loop}
\newacronym{fifo}{FIFO}{First-In-First-Out Buffer}

\newacronym{psc}{PSC}{post-synaptic current}
\newacronym{psp}{PSP}{post-synaptic potential}
\newacronym{stp}{STP}{short-term plasticity}
\newacronym{stdp}{STDP}{spike-timing-dependent plasticity}
\newacronym{rstdp}{R-STDP}{reward-modulated spike-timing-dependent plasticity}
\newacronym{lif}{LIF}{leaky integrate-and-fire}
\newacronym{adex}{AdEx}{adaptive exponential leaky integrate-and-fire}

\newacronym{mc}{MC}{Monte Carlo}

\newcommand{\affilKIP}{\textsuperscript{1}}
\newcommand{\affilDP}{\textsuperscript{2}}
\newcommand{\affilBIH}{\textsuperscript{3}}

\newcommand{\authorFIRST}{\IEEEauthorrefmark{1}}
\newcommand{\authorSENIOR}{\IEEEauthorrefmark{4}}

\begin{document}
\bstctlcite{IEEEexample:BSTcontrol}

\title{Versatile emulation of spiking neural networks on an accelerated neuromorphic substrate\vspace{-22pt}}

\author{
    \IEEEauthorblockN{\footnotesize
        S. Billaudelle\authorFIRST\affilKIP,
        Y. Stradmann\authorFIRST\affilKIP,
        K. Schreiber\authorFIRST\affilKIP,
        B. Cramer\authorFIRST\affilKIP,
        A. Baumbach\authorFIRST\affilKIP,
        D. Dold\authorFIRST\affilKIP,
	J. G{\"o}ltz\authorFIRST\affilKIP,
        A. F. Kungl\authorFIRST\affilKIP,
        T. C. Wunderlich\authorFIRST\affilBIH\affilKIP,
        A. Hartel\affilKIP,\\
        E. M{\"u}ller\affilKIP,
        O. Breitwieser\affilKIP,
        C. Mauch\affilKIP,
        M. Kleider\affilKIP,
        A. Gr{\"u}bl\affilKIP,
        D. St{\"o}ckel\affilKIP,
        C. Pehle\affilKIP,
        A. Heimbrecht\affilKIP,
        P. Spilger\affilKIP,
        G. Kiene\affilKIP,
        V. Karasenko\affilKIP,
        W. Senn\affilDP,\\
        M. A. Petrovici\authorFIRST\authorSENIOR\affilDP\affilKIP,
        J. Schemmel\authorSENIOR\affilKIP
        and K. Meier\authorSENIOR\affilKIP\IEEEauthorrefmark{2}
        }
    \vspace{5pt}
    \IEEEauthorblockA{\footnotesize
        \authorFIRST
        Authors with equal contribution \hspace{15pt}
        \authorSENIOR
        Shared senior authorship \hspace{15pt}
        \IEEEauthorrefmark{2}
        Deceased
        }
    \IEEEauthorblockA{\footnotesize
        \affilKIP
        Kirchhoff-Institute for Physics, Heidelberg University 
        \hspace{15pt}
        \affilDP
        Department of Physiology, University of Bern 
        }
    \IEEEauthorblockA{\footnotesize
        \affilBIH
        Berlin Institute of Health, Berlin / Charit\'{e}-Universitätsmedizin, Berlin
        }
    }


\maketitle

\vspace{-30pt}


\begin{abstract}
	We present first experimental results on the novel
BrainScaleS-2 neuromorphic architecture based on an analog neuro-synaptic core and augmented by embedded microprocessors for complex plasticity and experiment control.
The high acceleration factor of 1000 compared to biological dynamics enables the execution of computationally expensive tasks, by allowing the fast emulation of long-duration experiments or rapid iteration over many consecutive trials.
The flexibility of our architecture is demonstrated in a suite of five distinct experiments, which emphasize different aspects of the BrainScaleS-2 system.
\end{abstract}


%
\IEEEpeerreviewmaketitle

\section{Introduction}
\label{sec:intro}
The unifying principle behind all neuromorphic architectures lies in their attempt to emulate certain structural and dynamical aspects of biological nervous systems in order to inherit some of their well-known functional and metabolic
advantages over conventional silicon substrates.
However, precisely what these aspects are remains a holy grail of computational neuroscience, and our best attempts at answering this question are still rather conjectural.
This state of active exploration is reflected by the broad diversity of the current neuromorphic landscape \cite{thakur2018large}.

Consequently, our approach to neuromorphic engineering is explicitly geared towards building systems that can serve as scientific tools for studying this question.
By adhering to a restricted set of biologically inspired principles, we enable an efficient implementation in silico with respect to both emulation speed and power consumption.
Additionally, our proposed architecture emphasizes precision, scalability and, in particular, a substantial degree of flexibility.
This relates not only to the wide-ranged configurability of neuro-synaptic parameters and connectivity, but most importantly to the ability of influencing our circuits during emulation, which goes significantly beyond synaptic plasticity, as outlined below.

In this manuscript, we describe the core principles underlying the  BrainScaleS-2 architecture, followed by the emulation of a diverse set of spiking neural networks, which we have chosen to emphasize different aspects of the chip's operation and capabilities, as well as different computational principles that we believe are relevant for biological information processing.

\section{Architecture}
\label{seq:arch}
\begin{figure}[b]
	\begin{minipage}[t]{0.48\columnwidth}
		\vskip0pt
		\centering
		\includegraphics[width=0.90\textwidth]{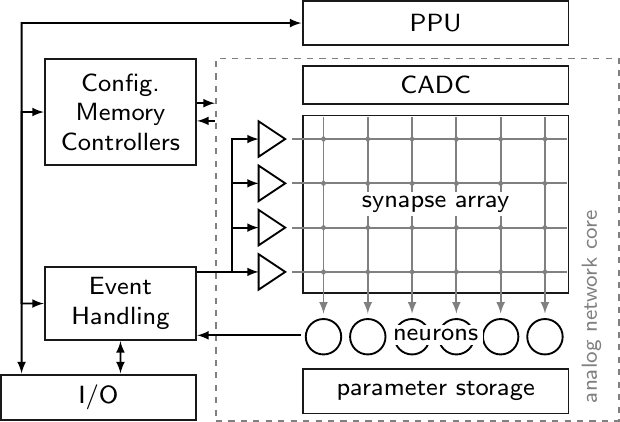}
	\end{minipage}\hfill
	\begin{minipage}[t]{0.48\columnwidth}
		\vspace{-0.17cm}
		\caption{
		    \textbf{Simplified block-level schematic of a \mbox{BrainScaleS-2} ASIC.}
		    The analog neuromorphic core is surrounded by event transport logic and control logic, including controllers for full-custom configuration SRAM.
		    Details and components that lie beyond the scope of this paper were omitted.
		    }
		\label{fig:arch}
	\end{minipage}
\end{figure}

BrainScaleS-2 is a family of mixed-signal neuromorphic systems implemented in a \SI{65}{\nano\meter} node.
It is centered around an analog neural network core implementing biologically inspired neuron and synapse circuits (\cref{fig:arch}).

The neurons \cite{aamir2018lifarray} feature \gls{lif} dynamics
with synaptic currents modeled as superpositions of spike-triggered exponential kernels.
The membrane is connected by a programmable conductance to a reset potential for a finite refractory period as soon as it crosses a certain threshold.
Additional mechanisms such as neuronal adaptation and exponential near-threshold dynamics \cite{aamir2018adex} or dendritic interaction \cite{schemmel2017accelerated} enable the emulation of more complex structure and dynamics.
All neurons are individually configurable via an on-chip analog parameter memory \cite{hock13analogmemory} and a set of digital control values.

Voltages and currents are -- scaled to utilize the available dynamic range -- directly represented in the respective circuits and evolve in continuous time.
Leveraging the intrinsic capacitances and conductances of the technology, time constants of neuron and synapse dynamics are rendered 1000 times smaller compared to typical values found in biology.
This thousandfold acceleration facilitates the execution of time-consuming tasks, such as performing high-dimensional parameter sweeps, the investigation of learning and metalearning, or statistical computations requiring large volumes of data \cite{cramer2019control, bohnstingl2019neuromorphic}.

Each neuron is associated with a column of synapse circuits \cite{friedmann2016hybridlearning}, which receive their inputs from the chip's digital backend.
The \SI{6}{\bit} synaptic weight is stored in local \gls{sram}. It further holds a \SI{6}{\bit} label, enabling synapses to filter afferent events tagged with their respective source address.
Each synapse also implements an analog circuit for measuring pairwise correlations between pre- and post-synaptic spike events \cite{friedmann2016hybridlearning}, enabling access to various forms of \gls{stdp}.
The analog correlation traces are accessible via \glspl{cadc}, which also allow the digitization of neuronal observables such as the membrane potential.

The versatility of the BrainScaleS-2 architecture is substantially augmented by the incorporation of freely programmable embedded microprocessors \cite{friedmann2016hybridlearning}. 
Together with their \gls{simd} vector units, which are tightly coupled to the synapse arrays' \gls{sram} controllers and the \glspl{cadc}, they form \glspl{ppu} for efficient control of synaptic plasticity.
Access to the on-chip configuration bus further allows the processor to also reconfigure all other components of the neuromorphic system during experiment execution.
The \gls{ppu} can thus be used for a vast array of applications such as near-arbitrary learning rules, on-line circuit calibration, structural network reconfiguration, or the co-simulation of an environment capable of continuous interaction with the network running on the neuromorphic core.

The experiments below were implemented on two revisions of the BrainScaleS-2 architecture.
Sections~\ref{sub:sampling} through~\ref{sub:insects} cover experiments conducted on a prototype featuring 32 neurons and 32$\times$32 synapses.
A full-size system with 512 neurons and 512$\times$256 synapses was used for \ref{sub:ttfs}.

\section{Experiments}


\subsection{Deep learning using precise spike timing}
\label{sub:ttfs} 
In many applications, the time and energy to solution represent essential commodities.
For spiking networks,  optimal use of these resources often imposes to have as few and as early spikes as possible.
However, the discrete nature of spikes makes it difficult to apply conventional machine learning algorithms based on differentiable loss functions.

\begin{figure}
    \centering
	\includegraphics[width=0.48\textwidth]{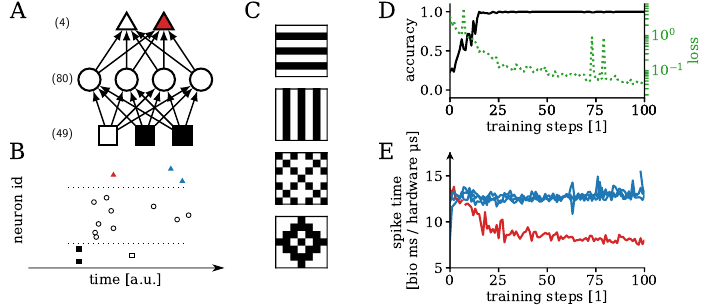}
    \caption{
	\textbf{Pattern recognition with time-to-first-spike coding.}
		\textbf{A)}
		Hierarchical network structure and neuron numbers per layer.
		\textbf{B)}
		Input ({\scriptsize $\square$}) spike times encode input pixel brightness.
		Activity propagates through the hidden layer ({\normalsize $\circ$}) to the label layer ($\triangle$).
		There, classification is determined by the identity of the first neuron to spike (red).
		\textbf{C)}
		Training/test set consisting of four patterns.
		\textbf{D)}
		Accuracy increase and corresponding decrease of loss during learning.
		\textbf{E)}
		Evolution of label neuron spike times during training for one example image.
	}	
	\label{fig:ttfsPlot}
\end{figure}

In the time-to-first-spike coding scheme, a neuron encodes a continuous variable as the time elapsed before its first spike.
The decision of a network performing a classification task is given by the first neuron to spike in the label layer (\cref{fig:ttfsPlot}A,B).
For such networks, an efficient gradient-descent-based learning scheme was first proposed in \cite{mostafa2017supervised}, using error backpropagation on a continuous function of output spike times.

We have generalized this method to include an exact, closed-form expression for finite membrane time constants \cite{goeltz2019mastersthesis,goeltz2019fast} and applied it to a 3-layer network emulated on \mbox{BrainScaleS-2} (\cref{fig:ttfsPlot}).
The loss was calculated as the cross-entropy of a softmax function on negative spike times, in order to maximize the distance between correct and incorrect label layer spikes.
\cref{fig:ttfsPlot}D shows its evolution during the host-based training and the associated classification accuracy for a simple 4-class learning task (\cref{fig:ttfsPlot}C).
The evolution of the label neuron spike times for one example class is shown in \cref{fig:ttfsPlot}E.
The robustness of the applied learning rule and the emulated network dynamics is evidenced by the clear separation of first-spike times.

A particularly appealing feature of this implementation is its extreme communication sparsity, with only one input spike per input variable and at most one spike per emulated neuron before classification.
After learning, the emulated network needed less than $\SI{10}{\micro\second}$ to classify an image.
This duration scales proportionally to the chosen synaptic and membrane time constants, which in our case were set to $\SI{5}{\micro\second}$.
Taking into consideration relaxation times between patterns, our setup is able to handle a pattern throughput of at least $\SI{20}{\kilo\Hz}$, independently of emulated network size~\cite{goeltz2019fast}.

\subsection{Sampling-based Bayesian computation}
\label{sub:sampling}
The Bayesian brain hypothesis \cite{doya2007bayesian} aspires to explain how the mammalian brain can operate in a probabilistic sea of sensory data.
In \cite{petrovici2016stochastic}, it was shown how networks of LIF neurons can learn to perform Bayesian inference through sampling on high-dimensional data distributions \cite{leng2018spiking, dold2019stochasticity}.

\begin{figure}
    \centering
    \includegraphics[width=0.48\textwidth]{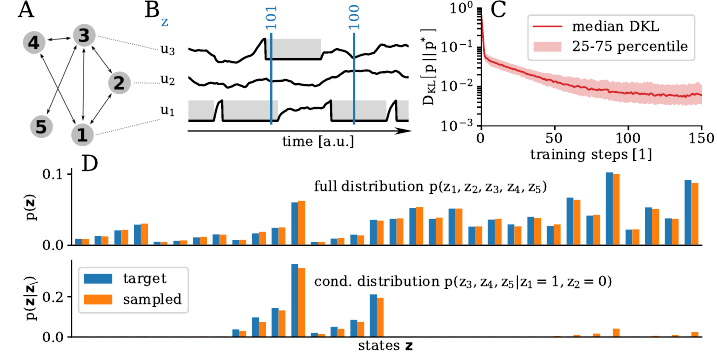}
    \caption{\textbf{Spike-based Bayesian inference.}
    \textbf{A)} Schematic of a random spiking sampling network.
    \textbf{B)} Membrane voltages of three selected neurons and visualisation of the spike-based representation of binary random variables.
    \textbf{C)} Sampling performance after training for 500 randomly generated target distributions.
    \textbf{D)} Sampling from the learned (top) and an associated conditional distribution (bottom).
                Orange: sampled distribution.
                Blue: analytically calculated target distribution.
                Remaining error bars are too small to visualize.
    }
    \label{fig:samplingPlot}
\end{figure}

In this quintessentially spike-based framework, neurons become stochastic due to background spiking input, thereby lending themselves to the representation of binary random variables: during post-spike refractoriness, a neuron is considered to be in the state $z=1$, and $z=0$ otherwise (\cref{fig:samplingPlot}A,B).
With appropriate synaptic connections, the resulting network dynamics inherently generate a sequence of samples from the learned distribution.
This enables the training of spiking networks to perform sampling-based Bayesian inference in arbitrary binary probability spaces, with applications to generative as well as discriminative problems \cite{probst2015probabilistic, kungl2018generative}.

Contrastive Hebbian learning \cite{hinton1984boltzmann} was performed with the hardware in the loop, i.e., with updates calculated on a host PC \cite{schmitt2017neuromorphic,kungl2018generative}.
Each training step was run for \SI{100}{\milli\second}, corresponding to \SI{100}{s} bio time and approximately \num{5000} samples.

Training was monitored using the Kullback-Leibler divergence between sampled and target distribution (\cref{fig:samplingPlot}C).
After training, the network reliably sampled from its target distribution and from associated conditional distributions (Bayesian inference, \cref{fig:samplingPlot}D).
Compared to previous neuromorphic realizations of neural sampling with analog neurons \cite{petrovici2017pattern,kungl2018generative}, the BrainScaleS-2 system allows unprecedented precision, while still enabling fast inference due to its thousandfold acceleration.

\subsection{Reinforcement learning}
\label{sub:pong}
Recent advances in reinforcement learning have enabled artificial systems to achieve unprecedented performance in board and computer games \cite{sutton2018reinforcement}.
Having clear roots in neurobiology, it is also an interesting framework for neuromorphic agents that learn through repeated interaction with an environment.

\begin{figure}
    \centering
    \includegraphics[width=0.48\textwidth]{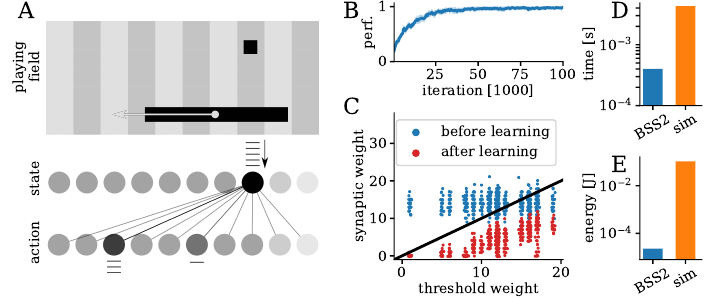}
    \caption{
        \textbf{Reinforcement learning with reward-modulated STDP.}
        \textbf{A)} The \gls{ppu} simulates a simplified version of Pong.
                    The horizontal position of the ball serves as input for a 2-layer neural network, with the resulting output dictating the target paddle position.
                    The network receives reward based on its aiming accuracy.
        \textbf{B)} Playing performance
                    during learning.
        \textbf{C)} Synaptic depression automatically adapts to the excitability of neurons.
        \textbf{D, E)} Wall-clock duration and power consumption of a single iteration on BrainScaleS-2 (blue) and an equivalent software simulation using NEST (orange).
        }
    \label{fig:pong}
\end{figure}

Three-factor learning rules \cite{fremaux2016neuromodulated} can implement reinforcement learning in spiking neural networks using a global neuromodulator and local observables such as spike rates.
As already shown in \cite{wunderlich2019demonstrating}, the BrainScaleS-2 architecture supports the implementation of an R-STDP learning rule \cite{fremaux2010functional,fremaux2016neuromodulated} in a closed-loop setup contained fully on chip.
Its application to a simplified version of the Pong video game is shown in \cref{fig:pong}A.
The network dynamics were emulated by the neuromorphic substrate, while the embedded plasticity processor took on a dual role.
First, it simulated the game dynamics, creating a host-independent setup.
Second, it calculated the plasticity updates using the synaptically stored correlation traces according to  $\Delta w_{ij} \propto \left( R - \bar R\right) e_{ij}$, where $R$ is the reward, $\bar R$ its moving average and $e_{ij}$ an \gls{stdp}-like eligibility trace.

During training, the network learned to keep the ball close to the middle of the paddle (\cref{fig:pong}B).
Implicitly, the experiment also demonstrates how learning can compensate analog fixed-pattern noise (\cref{fig:pong}C): while the excitability of uncalibrated neurons varied significantly due to mismatch effects, synapses that would negatively impact correct tracking of the ball were systematically depressed to a subthreshold strength with respect to their postsynaptic neuron. 
Furthermore, this setup demonstrates the speed and power advantages of the BrainScaleS-2 architecture compared to software simulations (\cref{fig:pong}D/E).

\subsection{Structural plasticity}
\label{sub:structural}
Synaptic plasticity is well known to not only be limited to adjusting the strength of synapses; the connectome itself undergoes continuous change during the lifetime of an individual \cite{holtmaat2005transient,loewenstein2011multiplicative,kappel2015synaptic}.
By constraining the number of expressed synapses to enforce a certain level of sparsity, the nervous system appears to manage its spatial and energetic budget \cite{knoblauch2016structural}.
Similar constraints apply to all physical information-processing systems, with neuromorphic ones being no exception.
In particular, the synaptic fan-in of silicon neurons is often limited.

\begin{figure}
	\centering
	\includegraphics[width=0.48\textwidth]{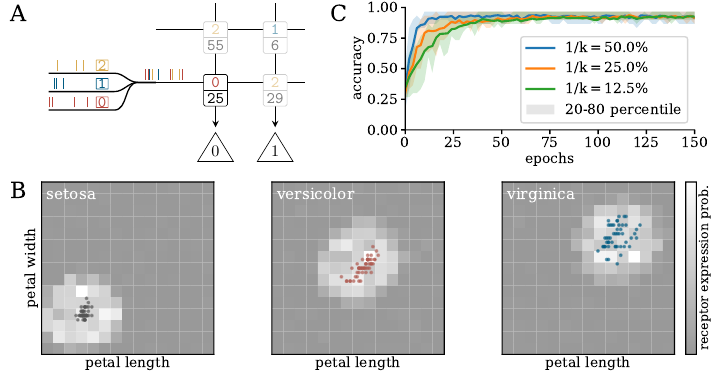}
	\caption{\textbf{Self-organizing receptive fields through structural plasticity.}
		\textbf{A)}
		Spike trains from different sources can be injected into a single synaptic row.
		Each synapse filters afferent spikes according to a locally stored label.
		\textbf{B)}
		A network endowed with structural plasticity learns to discriminate between types of Iris flowers (dataset represented by colored dots).
		The receptor distribution after training is adapted to the input data distribution.
		\textbf{C)}
		Feature selection through structural plasticity allows the conservation of classification performance even for strongly enforced sparsity $1-1/k$.
	}
	\label{fig:structural}
\end{figure}

We implemented a synaptic update policy that incorporates structural plasticity, enabling neurons to dynamically select a set of suitable synapses out of a pool of potential connections, that optimizes performance for a chosen task while maintaining a sparse connectome \cite{billaudelle2019structural}.
The learning rule is composed of three parts: an \gls{stdp} term that potentiates correlated connections, a homeostatic regularizer that limits post-synaptic firing rates and encourages synaptic competition, and a stochastic component that induces exploration.
A pruning condition is executed periodically, removing synapses with a weight below a certain threshold and randomly reassigning them.

Structural plasticity is enabled by bundling $k$ presynaptic sources and injecting them into a single synapse row (\cref{fig:structural}A): as each synapse can only gate one of these to its home neuron, pruning and reassigning of a synapse is simply implemented by changing its label.
The reconfiguration is thereby fully local and, in particular, does not involve time-consuming sorting of routing tables or connectivity lists~\cite{liu2018memory}.
If bundles are disjoint, their size $k$ also effectively sets the synaptic sparsity to $1 - 1/k$.

We applied the above algorithm to a supervised learning task, where the network was trained to classify the Iris data set \cite{fisher1936use}.
We randomly placed 48 receptor neurons on the two-dimensional feature plane spanned by petal width and length.
The firing rate of a receptor was set to increase with its proximity to a presented data point.
In three separate scenarios, the resulting $n=48$ input spike trains were injected into $m=6$, $12$, and $24$ synaptic rows, leading to three different levels of sparsity:
each label neuron could only see $1/k = m/n = \SI{12.5}{\%}$, $\SI{25.0}{\%}$, and $\SI{50.0}{\%}$ of the receptors at each point in time, respectively.
During training, teacher stimuli ensured that the correct label neurons were excited when an input belonging to their respective class was presented.

The emulated plasticity rule led to self-organized reconfiguration of their receptive fields (\cref{fig:structural}B), as the correlation between teacher signal and receptor proximity to the presented data drove the potentiation of associated synaptic weights.
For higher degrees of enforced sparsity, convergence times were longer, as the search for relevant inputs in the feature space became statistically more challenging.
Ultimately however, the learning rule enabled the network to achieve near-perfect classification in all three scenarios (\cref{fig:structural}C), demonstrating its ability to ensure a better utilization of synaptic resources without prior knowledge of the input data.


\subsection{Insect navigation}
\label{sub:insects}
\begin{figure}
    \centering
    \includegraphics[width=0.48\textwidth]{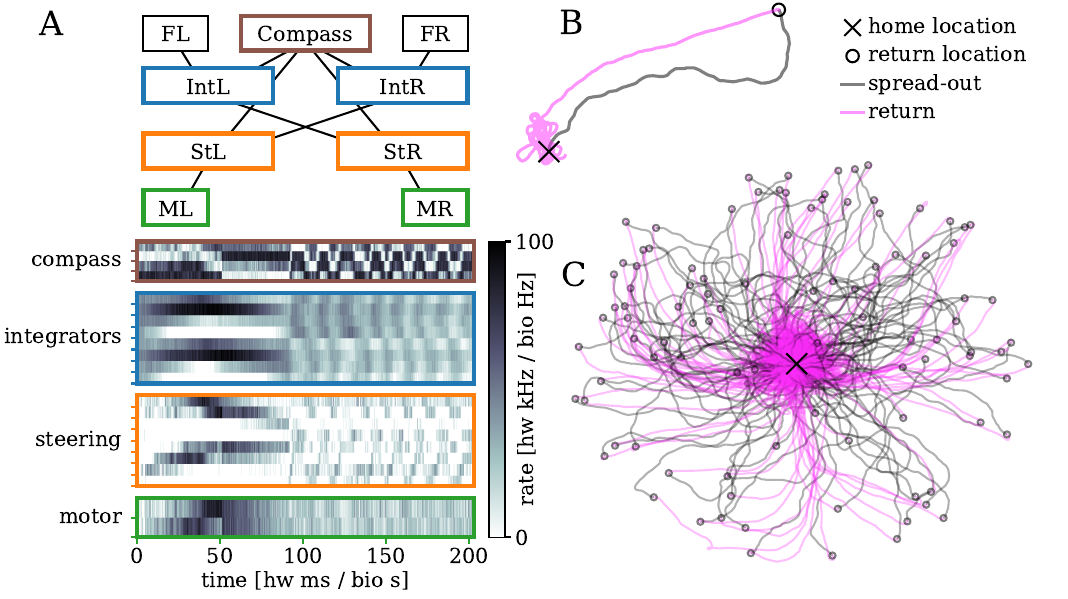}
    \caption{
        \textbf{Virtual insectoid agent performing path integration on BrainScaleS-2.}
        \textbf{A)} Network schematic and activity histogram.
                    The information flows from the sensory layer at the top through an integration and a steering layer to the motor neurons at the bottom.
                    R and L indicate the right and left side, respectively.
        \textbf{B)} A typical trajectory of the virtual insect which turns to random looping around the home position upon reaching it.
        \textbf{C)} Overlay of 100 trajectories like in B), each with a different random outbound journey.
    }
    \label{fig:insects}
\end{figure}

Recent developments in biological imaging and data processing have facilitated unprecedented insight into numerous functional aspects of insect brains \cite{chiang2011three, takemura2013visual, takemura2017connectome}.
For example, it has been shown that a structure known as the central complex is involved in navigational behavior~\cite{neuser2008analysis}. 
Based on physiological data from the bee's central complex and following \cite{stone2017anatomically}, we emulated a network for path integration (\cref{fig:insects}A) that reproduces bees' ability to return to their nest's location after exploring the environment for sources of food.

Each experiment started with a spread-out phase, in which a virtual insect performed a random walk starting from a certain origin.
During this phase, the modeled network received the sensory data of the absolute head orientation and the optical flow field of a left and right eye, but had no effect on the insect motion.
In the second part of the experiment, the return phase, the insect's motion was determined by two motor neurons within the network.
The insect's head orientation was encoded by four spike sources, each representing a cardinal direction similar to a compass.
The optical flow field was similarly represented by two spike generators that fired with a rate proportional to the optical flow as derived from the left and right eye, respectively (FL, FR).
Moreover, the two motor neurons (ML, MR) steered the insect by providing propulsion on the left or right hand side, similar to a tank drive.

While the model in \cite{stone2017anatomically} comprises 90 fire-rate-neurons with floating-point precision, the network on BrainScaleS-2 achieved about the same functional performance with only 18 neurons.
Additionally, we implemented the short-term memory mechanism employed by the integrator neurons to store directional distance as a synaptic mechanism.

The total flight duration was set to \SI{200}{ms} on the hardware, which corresponds to \SI{200}{s} in biology.
In that time, sensory information and steering signals were exchanged between body and brain every \SI{100}{\micro\second}.
During the first \SI{50}{\milli\second} the insect performed a random outbound journey, after which it returned to the nest.
Sample trajectories can be seen in \cref{fig:insects}B,C.
The average spike rate of all neurons and spike generators was \SI{30}{\kilo\Hz} (\SI{30}{\Hz} bio), which is in good agreement with experimental data from drosophila~\cite{gouwens2009signal} or locusts~\cite{moreaux2007estimating}.

In this experiment, the \gls{ppu} handled multiple tasks: the processing of synaptic modulations for the integrator neurons, the simulation of the environment, an emulation of all sensors including the corresponding spike stimuli, the translation of neuronal data into actions of motion, and the entire experiment control.
Apart from the setup and readout phase, the experiment ran entirely self-contained on the BrainScaleS-2 system.

\section{Discussion and outlook}
\label{sec:discussion}
In a post-Moore era, neuromorphic circuits represent a promising venue for advancing the computational capabilities of silicon.
This manuscript motivates how, by coupling the advantages of analog circuits with the flexibility of general-purpose digital computation and control, our BrainScaleS-2 architecture contributes to the research-oriented territory of the neuromorphic landscape.
In our endeavor, we share a common goal with other promising architectures such as \cite{furber2014spinnaker} and \cite{davies2018loihi}, which follow radically different design paradigms with advantages and drawbacks of their own.

A key aspect that we do not address above, but ultimately decides the value of such systems for computational neuroscience research, is their scalability.
Following integration concepts first proposed in \cite{schemmel2010wafer} and studied in, e.g., \cite{petrovici2014characterization, schmitt2017neuromorphic}, the BrainScaleS-2 architecture is explicitly designed to scale up to large, multi-chip systems.
These will conserve the network-size-independent speedup and energy efficiency that we have addressed in our above experiments, thus providing access to spiking network studies that are otherwise prohibitive for simulation software running on conventional substrates.


\section*{Acknowledgements}

We gratefully acknowledge funding from the European Union
under grant agreements 604102, 720270, 785907 (HBP) and the Manfred St{\"a}rk Foundation.


%

\bibliographystyle{IEEEtran}
\bibliography{bib}

\begin{thebibliography}{10}
\providecommand{\url}[1]{#1}
\csname url@samestyle\endcsname
\providecommand{\newblock}{\relax}
\providecommand{\bibinfo}[2]{#2}
\providecommand{\BIBentrySTDinterwordspacing}{\spaceskip=0pt\relax}
\providecommand{\BIBentryALTinterwordstretchfactor}{4}
\providecommand{\BIBentryALTinterwordspacing}{\spaceskip=\fontdimen2\font plus
\BIBentryALTinterwordstretchfactor\fontdimen3\font minus
  \fontdimen4\font\relax}
\providecommand{\BIBforeignlanguage}[2]{{%
\expandafter\ifx\csname l@#1\endcsname\relax
\typeout{** WARNING: IEEEtran.bst: No hyphenation pattern has been}%
\typeout{** loaded for the language `#1'. Using the pattern for}%
\typeout{** the default language instead.}%
\else
\language=\csname l@#1\endcsname
\fi
#2}}
\providecommand{\BIBdecl}{\relax}
\BIBdecl

\bibitem{thakur2018large}
C.~S.~T. Thakur, J.~Molin, G.~Cauwenberghs, G.~Indiveri, K.~Kumar, N.~Qiao,
  J.~Schemmel, R.~M. Wang \emph{et~al.}, ``Large-scale neuromorphic spiking
  array processors: A quest to mimic the brain,'' \emph{Frontiers in
  neuroscience}, vol.~12, p. 891, 2018.

\bibitem{aamir2018lifarray}
S.~A. {Aamir}, Y.~{Stradmann}, P.~{Müller}, C.~{Pehle}, A.~{Hartel},
  A.~{Grübl}, J.~{Schemmel}, and K.~{Meier}, ``An accelerated lif neuronal
  network array for a large-scale mixed-signal neuromorphic architecture,''
  \emph{IEEE Transactions on Circuits and Systems I: Regular Papers}, vol.~65,
  no.~12, pp. 4299--4312, Dec 2018.

\bibitem{aamir2018adex}
S.~A. {Aamir}, P.~{Müller}, G.~{Kiene}, L.~{Kriener}, Y.~{Stradmann},
  A.~{Grübl}, J.~{Schemmel}, and K.~{Meier}, ``A mixed-signal structured adex
  neuron for accelerated neuromorphic cores,'' \emph{IEEE Transactions on
  Biomedical Circuits and Systems}, vol.~12, no.~5, pp. 1027--1037, Oct 2018.

\bibitem{schemmel2017accelerated}
J.~Schemmel, L.~Kriener, P.~M{\"u}ller, and K.~Meier, ``An accelerated analog
  neuromorphic hardware system emulating nmda-and calcium-based non-linear
  dendrites,'' in \emph{2017 International Joint Conference on Neural Networks
  (IJCNN)}.\hskip 1em plus 0.5em minus 0.4em\relax IEEE, 2017, pp. 2217--2226.

\bibitem{hock13analogmemory}
M.~Hock, A.~Hartel, J.~Schemmel, and K.~Meier, ``An analog dynamic memory array
  for neuromorphic hardware,'' in \emph{Circuit Theory and Design (ECCTD), 2013
  European Conference on}, Sep. 2013, pp. 1--4.

\bibitem{cramer2019control}
B.~Cramer, D.~St{\"o}ckel, M.~Kreft, J.~Schemmel, K.~Meier, and V.~Priesemann,
  ``Control of criticality and computation in spiking neuromorphic networks
  with plasticity,'' \emph{arXiv preprint arXiv:1909.08418}, 2019.

\bibitem{bohnstingl2019neuromorphic}
T.~Bohnstingl, F.~Scherr, C.~Pehle, K.~Meier, and W.~Maass, ``Neuromorphic
  hardware learns to learn,'' \emph{Frontiers in neuroscience}, vol.~13, 2019.

\bibitem{friedmann2016hybridlearning}
S.~Friedmann, J.~Schemmel, A.~Gr\"ubl, A.~Hartel, M.~Hock, and K.~Meier,
  ``Demonstrating hybrid learning in a flexible neuromorphic hardware system,''
  \emph{IEEE Transactions on Biomedical Circuits and Systems}, vol.~11, no.~1,
  pp. 128--142, 2017.

\bibitem{mostafa2017supervised}
H.~Mostafa, ``Supervised learning based on temporal coding in spiking neural
  networks,'' \emph{IEEE transactions on neural networks and learning systems},
  vol.~29, no.~7, pp. 3227--3235, 2017.

\bibitem{goeltz2019mastersthesis}
\BIBentryALTinterwordspacing
J.~G\"oltz, ``Training deep networks with time-to-first-spike coding on the
  brainscales wafer-scale system,'' Masterarbeit, Universit\"at Heidelberg,
  April 2019. [Online]. Available:
  \url{http://www.kip.uni-heidelberg.de/Veroeffentlichungen/details.php?id=3909}
\BIBentrySTDinterwordspacing

\bibitem{goeltz2019fast}
J.~G{\"o}ltz, A.~Baumbach, S.~Billaudelle, O.~Breitwieser, D.~Dold, L.~Kriener,
  A.~F. Kungl, W.~Senn \emph{et~al.}, ``Fast and deep neuromorphic learning
  with time-to-first-spike coding,'' \emph{arXiv preprint arXiv:1912.11443},
  2019.

\bibitem{doya2007bayesian}
K.~Doya, S.~Ishii, A.~Pouget, and R.~P. Rao, \emph{Bayesian brain:
  Probabilistic approaches to neural coding}.\hskip 1em plus 0.5em minus
  0.4em\relax MIT press, 2007.

\bibitem{petrovici2016stochastic}
M.~A. Petrovici, J.~Bill, I.~Bytschok, J.~Schemmel, and K.~Meier, ``Stochastic
  inference with spiking neurons in the high-conductance state,''
  \emph{Physical Review E}, vol.~94, no.~4, p. 042312, 2016.

\bibitem{leng2018spiking}
L.~Leng, R.~Martel, O.~Breitwieser, I.~Bytschok, W.~Senn, J.~Schemmel,
  K.~Meier, and M.~A. Petrovici, ``Spiking neurons with short-term synaptic
  plasticity form superior generative networks,'' \emph{Scientific reports},
  vol.~8, no.~1, p. 10651, 2018.

\bibitem{dold2019stochasticity}
D.~Dold, I.~Bytschok, A.~F. Kungl, A.~Baumbach, O.~Breitwieser, W.~Senn,
  J.~Schemmel, K.~Meier, and M.~A. Petrovici, ``Stochasticity from
  function—why the bayesian brain may need no noise,'' \emph{Neural
  Networks}, vol. 119, pp. 200--213, 2019.

\bibitem{probst2015probabilistic}
D.~Probst, M.~A. Petrovici, I.~Bytschok, J.~Bill, D.~Pecevski, J.~Schemmel, and
  K.~Meier, ``Probabilistic inference in discrete spaces can be implemented
  into networks of lif neurons,'' \emph{Frontiers in computational
  neuroscience}, vol.~9, p.~13, 2015.

\bibitem{kungl2018generative}
A.~F. Kungl, S.~Schmitt, J.~Kl{\"a}hn, P.~M{\"u}ller, A.~Baumbach, D.~Dold,
  A.~Kugele, N.~G{\"u}rtler \emph{et~al.}, ``Generative models on accelerated
  neuromorphic hardware,'' \emph{arXiv preprint arXiv:1807.02389}, 2018.

\bibitem{hinton1984boltzmann}
G.~E. Hinton, T.~J. Sejnowski, and D.~H. Ackley, \emph{Boltzmann machines:
  Constraint satisfaction networks that learn}.\hskip 1em plus 0.5em minus
  0.4em\relax Carnegie-Mellon University, Department of Computer Science
  Pittsburgh, 1984.

\bibitem{schmitt2017neuromorphic}
S.~Schmitt, J.~Kl{\"a}hn, G.~Bellec, A.~Gr{\"u}bl, M.~G{\"u}ttler, A.~Hartel,
  S.~Hartmann, D.~Husmann \emph{et~al.}, ``Neuromorphic hardware in the loop:
  Training a deep spiking network on the brainscales wafer-scale system,'' in
  \emph{2017 International Joint Conference on Neural Networks (IJCNN)}.\hskip
  1em plus 0.5em minus 0.4em\relax IEEE, 2017, pp. 2227--2234.

\bibitem{petrovici2017pattern}
M.~A. Petrovici, S.~Schmitt, J.~Kl{\"a}hn, D.~St{\"o}ckel, A.~Schroeder,
  G.~Bellec, J.~Bill, O.~Breitwieser \emph{et~al.}, ``Pattern representation
  and recognition with accelerated analog neuromorphic systems,'' in \emph{2017
  IEEE International Symposium on Circuits and Systems (ISCAS)}.\hskip 1em plus
  0.5em minus 0.4em\relax IEEE, 2017, pp. 1--4.

\bibitem{sutton2018reinforcement}
R.~S. Sutton and A.~G. Barto, \emph{Reinforcement learning: An
  introduction}.\hskip 1em plus 0.5em minus 0.4em\relax MIT press, 2018.

\bibitem{fremaux2016neuromodulated}
N.~Fr{\'e}maux and W.~Gerstner, ``Neuromodulated spike-timing-dependent
  plasticity, and theory of three-factor learning rules,'' \emph{Frontiers in
  neural circuits}, vol.~9, p.~85, 2016.

\bibitem{wunderlich2019demonstrating}
T.~Wunderlich, A.~F. Kungl, E.~M{\"u}ller, A.~Hartel, Y.~Stradmann, S.~A.
  Aamir, A.~Gr{\"u}bl, A.~Heimbrecht \emph{et~al.}, ``Demonstrating advantages
  of neuromorphic computation: a pilot study,'' \emph{Frontiers in
  Neuroscience}, vol.~13, p. 260, 2019.

\bibitem{fremaux2010functional}
N.~Fr{\'e}maux, H.~Sprekeler, and W.~Gerstner, ``Functional requirements for
  reward-modulated spike-timing-dependent plasticity,'' \emph{Journal of
  Neuroscience}, vol.~30, no.~40, pp. 13\,326--13\,337, 2010.

\bibitem{holtmaat2005transient}
A.~J. Holtmaat, J.~T. Trachtenberg, L.~Wilbrecht, G.~M. Shepherd, X.~Zhang,
  G.~W. Knott, and K.~Svoboda, ``Transient and persistent dendritic spines in
  the neocortex in vivo,'' \emph{Neuron}, vol.~45, no.~2, pp. 279--291, 2005.

\bibitem{loewenstein2011multiplicative}
Y.~Loewenstein, A.~Kuras, and S.~Rumpel, ``Multiplicative dynamics underlie the
  emergence of the log-normal distribution of spine sizes in the neocortex in
  vivo,'' \emph{Journal of Neuroscience}, vol.~31, no.~26, pp. 9481--9488,
  2011.

\bibitem{kappel2015synaptic}
D.~Kappel, S.~Habenschuss, R.~Legenstein, and W.~Maass, ``Synaptic sampling: A
  bayesian approach to neural network plasticity and rewiring,'' in
  \emph{Advances in Neural Information Processing Systems}, 2015, pp. 370--378.

\bibitem{knoblauch2016structural}
A.~Knoblauch and F.~T. Sommer, ``Structural plasticity, effectual connectivity,
  and memory in cortex,'' \emph{Frontiers in neuroanatomy}, vol.~10, p.~63,
  2016.

\bibitem{billaudelle2019structural}
S.~Billaudelle, B.~Cramer, M.~A. Petrovici, K.~Schreiber, D.~Kappel,
  J.~Schemmel, and K.~Meier, ``Structural plasticity on an accelerated analog
  neuromorphic hardware system,'' \emph{arXiv preprint arXiv:1912.12047}, 2019.

\bibitem{liu2018memory}
C.~Liu, G.~Bellec, B.~Vogginger, D.~Kappel, J.~Partzsch, F.~Neum{\"a}rker,
  S.~H{\"o}ppner, W.~Maass \emph{et~al.}, ``Memory-efficient deep learning on a
  spinnaker 2 prototype,'' \emph{Frontiers in neuroscience}, vol.~12, 2018.

\bibitem{fisher1936use}
R.~A. Fisher, ``The use of multiple measurements in taxonomic problems,''
  \emph{Annals of eugenics}, vol.~7, no.~2, pp. 179--188, 1936.

\bibitem{chiang2011three}
A.-S. Chiang, C.-Y. Lin, C.-C. Chuang, H.-M. Chang, C.-H. Hsieh, C.-W. Yeh,
  C.-T. Shih, J.-J. Wu \emph{et~al.}, ``Three-dimensional reconstruction of
  brain-wide wiring networks in drosophila at single-cell resolution,''
  \emph{Current Biology}, vol.~21, no.~1, pp. 1--11, 2011.

\bibitem{takemura2013visual}
S.~Takemura, A.~Bharioke, Z.~Lu, A.~Nern, S.~Vitaladevuni, P.~K. Rivlin, W.~T.
  Katz, D.~J. Olbris \emph{et~al.}, ``A visual motion detection circuit
  suggested by drosophila connectomics,'' \emph{Nature}, vol. 500, no. 7461, p.
  175, 2013.

\bibitem{takemura2017connectome}
S.-y. Takemura, Y.~Aso, T.~Hige, A.~Wong, Z.~Lu, C.~S. Xu, P.~K. Rivlin,
  H.~Hess \emph{et~al.}, ``A connectome of a learning and memory center in the
  adult drosophila brain,'' \emph{Elife}, vol.~6, p. e26975, 2017.

\bibitem{neuser2008analysis}
K.~Neuser, T.~Triphan, M.~Mronz, B.~Poeck, and R.~Strauss, ``Analysis of a
  spatial orientation memory in drosophila,'' \emph{Nature}, vol. 453, no.
  7199, p. 1244, 2008.

\bibitem{stone2017anatomically}
T.~Stone, B.~Webb, A.~Adden, N.~B. Weddig, A.~Honkanen, R.~Templin, W.~Wcislo,
  L.~Scimeca \emph{et~al.}, ``An anatomically constrained model for path
  integration in the bee brain,'' \emph{Current Biology}, vol.~27, no.~20, pp.
  3069--3085, 2017.

\bibitem{gouwens2009signal}
N.~W. Gouwens and R.~I. Wilson, ``Signal propagation in drosophila central
  neurons,'' \emph{Journal of Neuroscience}, vol.~29, no.~19, pp. 6239--6249,
  2009.

\bibitem{moreaux2007estimating}
L.~C. Moreaux and G.~Laurent, ``Estimating firing rates from calcium signals in
  locust projection neurons in vivo,'' \emph{Frontiers in neural circuits},
  vol.~1, p.~2, 2007.

\bibitem{furber2014spinnaker}
S.~B. Furber, F.~Galluppi, S.~Temple, and L.~A. Plana, ``The spinnaker
  project,'' \emph{Proceedings of the IEEE}, vol. 102, no.~5, pp. 652--665,
  2014.

\bibitem{davies2018loihi}
M.~Davies, N.~Srinivasa, T.-H. Lin, G.~Chinya, Y.~Cao, S.~H. Choday, G.~Dimou,
  P.~Joshi \emph{et~al.}, ``Loihi: A neuromorphic manycore processor with
  on-chip learning,'' \emph{IEEE Micro}, vol.~38, no.~1, pp. 82--99, 2018.

\bibitem{schemmel2010wafer}
J.~Schemmel, D.~Br{\"u}derle, A.~Gr{\"u}bl, M.~Hock, K.~Meier, and S.~Millner,
  ``A wafer-scale neuromorphic hardware system for large-scale neural
  modeling,'' in \emph{Proceedings of 2010 IEEE International Symposium on
  Circuits and Systems}.\hskip 1em plus 0.5em minus 0.4em\relax IEEE, 2010, pp.
  1947--1950.

\bibitem{petrovici2014characterization}
M.~A. Petrovici, B.~Vogginger, P.~M{\"u}ller, O.~Breitwieser, M.~Lundqvist,
  L.~Muller, M.~Ehrlich, A.~Destexhe \emph{et~al.}, ``Characterization and
  compensation of network-level anomalies in mixed-signal neuromorphic modeling
  platforms,'' \emph{PloS one}, vol.~9, no.~10, p. e108590, 2014.

\end{thebibliography}

%
%

\end{document}